\documentclass[aip,prl,graphicx,reprint]{revtex4-1}

\usepackage{graphicx}

\begin{document}


\title{High quality factor photonic cavity for dark matter axion searches} 



\author{D. Alesini}
\affiliation{INFN, Laboratori Nazionali di Frascati, Frascati (Roma) Italy}
\author{C. Braggio}
\affiliation{INFN, Sezione di Padova, Padova, Italy}
\affiliation{Dip.\,di Fisica e Astronomia,  Padova, Italy}
\author{G. Carugno}
\affiliation{INFN, Sezione di Padova, Padova, Italy}
\affiliation{Dip.\,di Fisica e Astronomia,  Padova, Italy}
\author{N. Crescini}
\affiliation{Dip.\,di Fisica e Astronomia,  Padova, Italy}
\affiliation{INFN, Laboratori Nazionali di Legnaro,  Legnaro (PD), Italy}
\author{D. D'\,Agostino}
\affiliation{Dip.\,di Fisica E.R. Caianiello, Fisciano (SA), Italy and INFN, Sez. di Napoli, Napoli, Italy}
\author{D. Di Gioacchino}
\affiliation{INFN, Laboratori Nazionali di Frascati, Frascati (Roma) Italy}
\author{R. Di Vora}
\affiliation{INFN, Sezione di Padova, Padova, Italy}
\affiliation{Dip. di Fisica, Siena, Italy}
\author{P. Falferi}
\affiliation{Istituto di Fotonica e Nanotecnologie, CNR, INFN - TIFPA and FBK,
Povo, Trento, Italy}
\author{U. Gambardella}
\affiliation{Dip.\,di Fisica E.R. Caianiello, Fisciano (SA), Italy and INFN, Sez. di Napoli, Napoli, Italy}
\author{C. Gatti}
\affiliation{INFN, Laboratori Nazionali di Frascati, Frascati (Roma) Italy}
\author{G. Iannone}
\affiliation{Dip.\,di Fisica E.R. Caianiello, Fisciano (SA), Italy and INFN, Sez. di Napoli, Napoli, Italy}
\author{C. Ligi}
\affiliation{INFN, Laboratori Nazionali di Frascati, Frascati (Roma) Italy}
\author{A. Lombardi}
\affiliation{INFN, Laboratori Nazionali di Legnaro,  Legnaro (PD), Italy}
\author{G. Maccarrone}
\affiliation{INFN, Laboratori Nazionali di Frascati, Frascati (Roma) Italy}
\author{A. Ortolan}
\affiliation{INFN, Laboratori Nazionali di Legnaro,  Legnaro (PD), Italy}
\author{R. Pengo}
\affiliation{INFN, Laboratori Nazionali di Legnaro,  Legnaro (PD), Italy}
\author{C. Pira}
\affiliation{INFN, Laboratori Nazionali di Legnaro,  Legnaro (PD), Italy}
\author{A. Rettaroli}
\affiliation{INFN, Laboratori Nazionali di Frascati, Frascati (Roma) Italy}
\affiliation{Dip.\,di Matematica e Fisica Universit\`a di Roma 3, Roma, Italy}
\author{G. Ruoso}
\email[Corresponding author ]{Giuseppe.Ruoso@lnl.infn.it}
\affiliation{INFN, Laboratori Nazionali di Legnaro,  Legnaro (PD), Italy}
\author{L. Taffarello}
\affiliation{INFN, Sezione di Padova, Padova, Italy}
\author{S. Tocci}
\affiliation{INFN, Laboratori Nazionali di Frascati, Frascati (Roma) Italy}

\collaboration{QUAX Collaboration}


\date{\today}

\begin{abstract}
Searches for dark matter axion involve the use of microwave resonant cavities operating in a strong magnetic field. Detector sensitivity is directly related to the cavity quality factor, which is limited, however, by the presence of the external magnetic field. In this paper we present a cavity of novel design whose quality factor is not affected by a magnetic field. It is based on a photonic structure by the use of  sapphire rods. The quality factor at cryogenic temperature is in excess of $5 \times 10^5$ for a selected mode. 

\end{abstract}


\maketitle 

\section{Introduction}

A remarkable  result of modern cosmology is that a major fraction of the mass content of the universe is composed of dark matter (DM), i.e. particles not interacting significantly with electromagnetic radiation, ordinary matter, not even self-interacting (cold dark matter) \cite{zwicky,rubin78,rubin80}. A class of particles generically indicated as Weakly Interacting Sub-eV Particles (WISP) recently came into attention as possible candidates as main DM component. WISPs  are very light bosons characterized by sub-eV masses and large occupation numbers, such as axions, axion-like particles (ALP) and hidden photons. The axion is the pseudo-Goldstone boson associated to an additional symmetry of the Standard Model Lagrangian which is spontaneously broken at an extremely high energy scale $F_a$ \cite{pq,weinberg1978new,wilczek1978problem,PhysRevLett.43.103,SHIFMAN1980493,DINE1981199,Zhitnitsky:1980tq}. For scales $F_a  =10^{12} $ GeV, corresponding to typical mass values $m_a =1 $ meV, axions may account for the totality of DM. As a consequence, many different detectors have been proposed over the last decades to search for relic axions \cite{Irastorza:2018dyq}. Although theory do not fix the value of $F_a$, cosmological considerations and astrophysical observations provide boundaries on $F_a$ and suggest a favoured axion mass range 1 $\mu$eV  $ < m_a < $ 10 meV.
Most of axion detectors rely on the conversion of axions into photons in a resonant cavity in the presence of a static magnetic field, following the detection scheme proposed by P. Sikivie in 1983  called axion haloscope, which is based on the inverse Primakoff effect \cite{PhysRevLett.51.1415}. After the pioneering work of the Rochester-Brookhaven-Fermilab collaboration \cite{DePanfilis:1987dk, Wuensch:1989sa} and of the group located at the University of Florida \cite{Hagmann:1990tj}, the Axion Dark Matter eXperiment (ADMX) collaboration operated for over two decades the most advanced detector of this type.
In particular, the ADMX experiment reached the cosmologically relevant sensitivity to exclude the axion mass range 2.66 $ < m_a < $ 3.31 $\mu$eV for DFSZ models \cite{PhysRevLett.120.151301, braine2019extended} and 1.91 $ < m_a < $ 3.69 $\mu$eV for KSVZ models \cite{PhysRevD.64.092003,PhysRevLett.104.041301}. Very recently many other haloscopes came into play and some of them are already running and taking data: among them HAYSTAC\cite{Brubaker:2016ktl}, ORGAN\,\cite{MCALLISTER201767}, CAPP\cite{Chung:2018wms,semertzidis2019axion,lee2020axion}, KLASH\cite{Alesini:2017ifp,Gatti:2018ojx,alesini2019klash}, RADES\cite{Melc_n_2018} and QUAX-a$\gamma$ \cite{PhysRevD.99.101101,DiGioacchino:2019coj}. While all these are bases on the Sikivie's scheme, new techniques have also been proposed, like for example the dielectric haloscopes of MADMAX \cite{PhysRevLett.118.091801} and BRASS \cite{BRASS}. 

In a standard haloscope the detector is composed by a large quality factor $Q$ resonant cavity immersed in a static magnetic field $B_0$. 
When the resonant frequency of the cavity $\nu_c$ is tuned to the corresponding axion mass $m_ac^2/h$, the expected power deposited by DM axions is given by\,\cite{ALKENANY201711,Brubaker:2016ktl}
	\begin{equation}
	\label{eq:power}
	P_{a}=\left( g_{\gamma}^2\frac{\alpha^2}{\pi^2}\frac{\hbar^3 c^3\rho_a}{\Lambda^4} \right) \times
	\left( \frac{\beta}{(1+\beta)^2} \omega_c \frac{1}{\mu_0} B_0^2 V C_{mnl} Q_0 \right),
	\end{equation}
where $\rho_a=0.45$\,GeV/cm$^3$ is the local DM density, $\alpha$ is the fine-structure constant, $\Lambda=78$\,MeV is a scale parameter related to hadronic physics, and $g_{\gamma}$ is a model dependent parameter equal to $-0.97$ $(0.36)$ in the KSVZ (DFSZ) axion model\,\cite{DINE1981199,SHIFMAN1980493,PhysRevLett.43.103,DINE1983137}.
It is related to the coupling appearing in the Lagrangian $g_{a\gamma\gamma}=(g_{\gamma}\alpha/\pi\Lambda^2)m_a$. The second parentheses contains the vacuum permeability $\mu_0$, the magnetic field strength $B_0$, the cavity volume $V$, its angular frequency $\omega_c=2\pi\nu_c$, the coupling between cavity and receiver $\beta$ and the unloaded  quality factor $Q_0$. $C_{mnl}\simeq O(1)$ is a geometrical factor, depending on the cavity mode, defined as the overlap between the microwave electric field $\mathbf{E}_{mnl}$ and the static magnetic field~$\mathbf{ B}$
\begin{equation}
C_{mnl} = \frac{\left[\int_V dV \mathbf{E}_{mnl} \cdot \mathbf{ B}\right]^2}
{V B^2 \int_V dV \epsilon_r E^2_{mnl}}
\end{equation}
A typical cavity geometry is of cylindrical type, which easily fits inside a superconducting solenoid providing a strong magnetic field. The cavity mode TM010 exhibits the largest coupling $C_{010} = 0.69$. This mode corresponds to a field distribution with maximum value of the electric field along the cavity axis.

The axion mass range studied by haloscopes up to now is limited to few $\mu$eV.
Exploring larger ranges at higher values requires the excitation of modes with frequency above a few GHz where several experimental limitations occur:
(\textit{i}) the available linear amplifiers limit the sensitivity\cite{Lamoreaux:2013koa}; (\textit{ii}) conversion volumes are smaller since the normal modes resonant frequencies are inversely proportional to the cavity radius; and (\textit{iii}) the anomalous skin effect reduces the  quality factor of copper cavities at high frequencies. Solutions to the first two issues are proposed for instance in\,\cite{Kuzmin:2018eeb} and\,\cite{Jeong:2017hqs}, respectively.

The optimum value of quality factor for haloscopes is  $\sim10^6$, as estimated by the coherence time of DM axions\,\cite{Barbieri:2016vwg}.
A 10\,GHz copper cavity, cooled at cryogenic temperature, barely reaches $Q\lesssim10^5$, a value that rapidly decreases with increasing frequency.  In the first phase of the experiment QUAX-a$\gamma$ \cite{PhysRevD.99.101101}, a solution to this problem has already been studied: a special type of hybrid superconducting cavity has been realized capable of operating inside a magnetic field with $Q_0$ in excess of $10^5$ for frequencies about 10 GHz. In this paper we  show that another viable possibility is the use of dielectric resonators, i.e. cavities in which a low-loss material with high dielectric coefficient is inserted to shape and concentrate the mode, reducing the dissipation on the cavity walls and consequently increasing the quality factor. These type of cavities are normally referred to as {\it photonic cavities}. This work as been conducted within the experiment QUAX-a$\gamma$.

\section{Design and Realization} 
\label{design}

Inspired by previous works~\cite{ISI:A1997WR98100025,ISI:000073912500031,ISI:000398869100003}, we designed a novel type of distributed Bragg reflector resonator working on the TM010-like mode at about 13.5 GHz. The choice of the frequency was mainly given by the possibility of incorporating the resonator in the detection chain developed for the QUAX experiment to search for dark matter axions~\cite{Crescini2018}.
The main idea was to design a pseudo-cylindrical cavity realizing an interference pattern to confine the TM010 through long dielectric rods placed parallel to the cavity axis. These types of cavities are also known as photonic band gap cavities~\cite{Kroll:1992xu,Woollett:2018htk}. The dielectrics, commercially available, were in the form of long sapphire cylinders of 2 mm diameter. Sapphire was chosen for its very low loss-tangent, going from about $10^{-5}$, at room temperature, down to a fraction of $10^{-7}$ at cryogenic temperatures~\cite{Krupka_1999}.

\begin{figure}[htbp]
  \begin{center}
    \includegraphics[width=.33\textwidth]{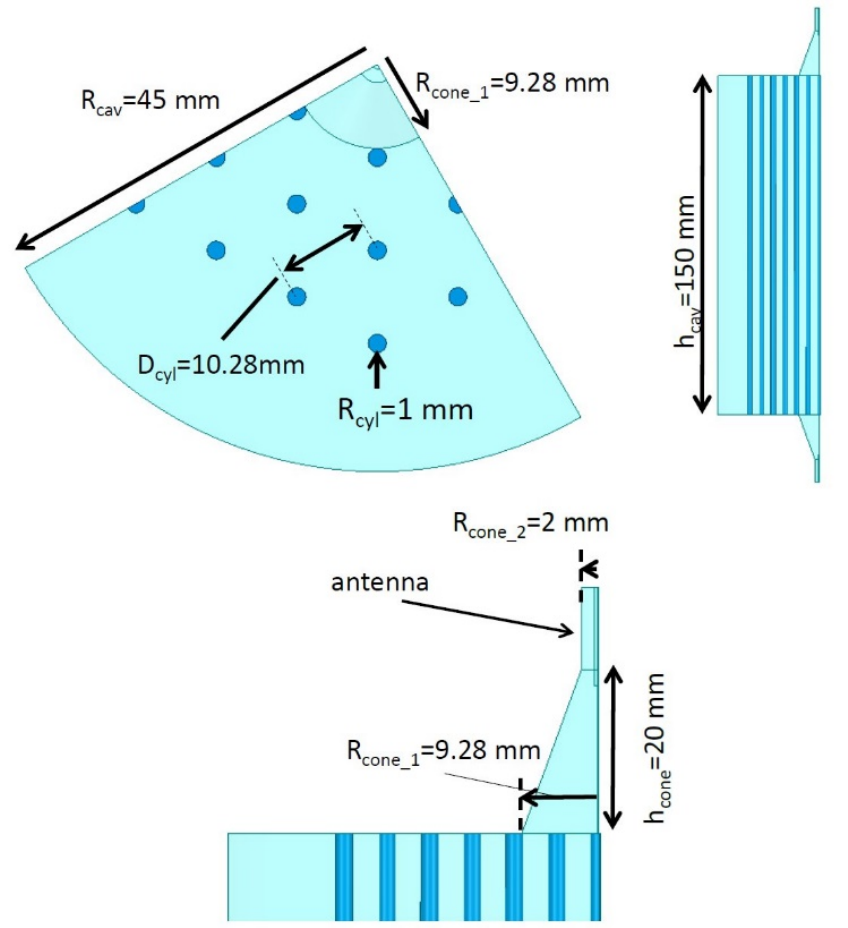}
    \caption{Final design of the  cavity with its relevant dimensions.}
    \label{fig1}
  \end{center}
\end{figure}

We did the electromagnetic calculations using the code ANSYS Electronics~\cite{Ansys}. Because of cylindrical symmetry
we simulated only one quarter of the structure with perfect magnetic boundary conditions.
The final designed cavity with relevant dimensions is given in Figure~\ref{fig1}.
In the cavity 36 sapphire rods of 2~mm diameter are positioned with the pattern shown in figure while the distance between the rods was chosen to have the resonant frequency of the confined mode TM010 at around 13.5~GHz.
The number of rods was chosen to have enough attenuation of the electromagnetic field toward the cavity outer wall.
Since the sapphire dielectric constant varies in a wide range depending on the crystal orientations, impurities and temperature,
in the simulations we have considered a sapphire dielectric constant equal to 11.1 and a loss tangent at cryogenic
temperature of $10^{-7}$ (See ~\cite{Krupka_1999}). The copper surface resistance was set to about 5.5 m$\Omega$ as expected in the anomalous regime  at this frequency~\cite{Reuter}. 
The two conical end plates were designed in order to reduce their loss contribution to a negligible level. In particular, as already implemented and illustrated in ref. \cite{PhysRevD.99.101101,DiGioacchino:2019coj}, the length of the conical shapes is such as to permit a sufficient attenuation of the electromagnetic field at the end plates.
To excite and detect the resonant modes, two coaxial antennas were inserted
at the end of the cones.

\begin{figure}[htbp]
  \begin{center}
    \includegraphics[width=.45\textwidth]{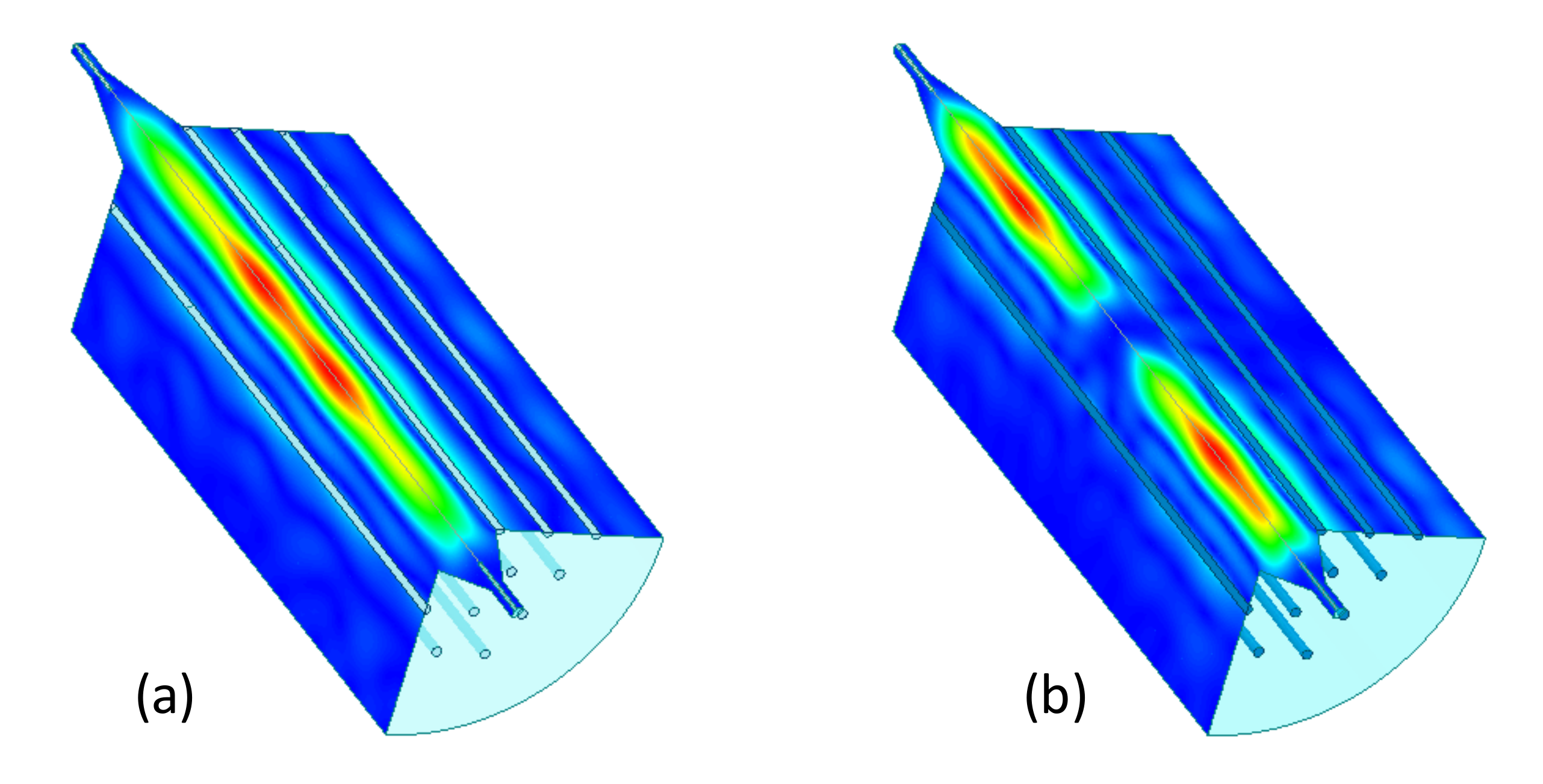}
    \caption{Magnitude of the electric field of the first two TM resonant modes: TM010 (a) and TM011 (b).}
    \label{fig2}
  \end{center}
\end{figure}

\begin{figure}[htbp]
  \begin{center}
   \includegraphics[width=.38\textwidth]{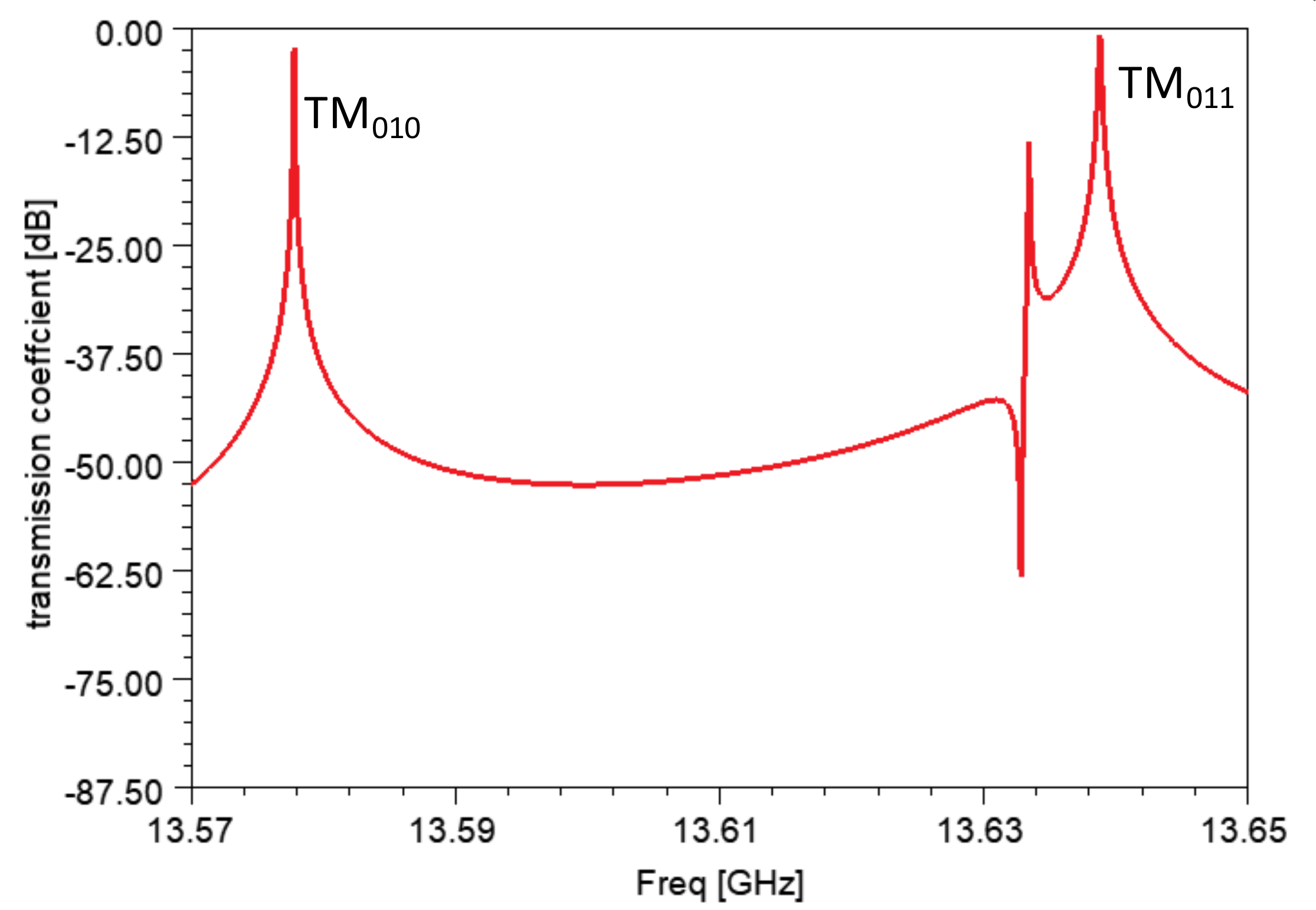}
    \caption{Simulated transmission coefficient between the two coupled antennas. The peaks correspond to the resonant modes TM010 and TM011. }
    \label{fig3}
  \end{center}
\end{figure}

The magnitude of the electric field of the first two TM resonant modes, TM010 and TM011, is given in Figure~\ref{fig2}.
The first mode is expected to have a quality factor of about $3\times10^5$ at liquid Helium temperature, essentially limited by the
losses on the cavity walls and endplates. Figure~\ref{fig3} shows the simulated transmission coefficient between the two
coupled antennas whose peaks correspond to the resonant modes TM010 and TM011, respectively. In the plot is also visible
one spurious peak corresponding to a mode configuration weakly coupled to the coaxial antennas.

By using the simulations it is easy to calculate the effective volume of the system, essential to deduce the axion signal. We got $C_{mnl} \times V = 0.024 \times 9.5 \times 10^{-4} =V_{\rm eff}^{\rm Sa} = 2.3 \cdot 10^{-5}$ m$^3$.

\begin{figure}[h]
\includegraphics[width=.42\textwidth]{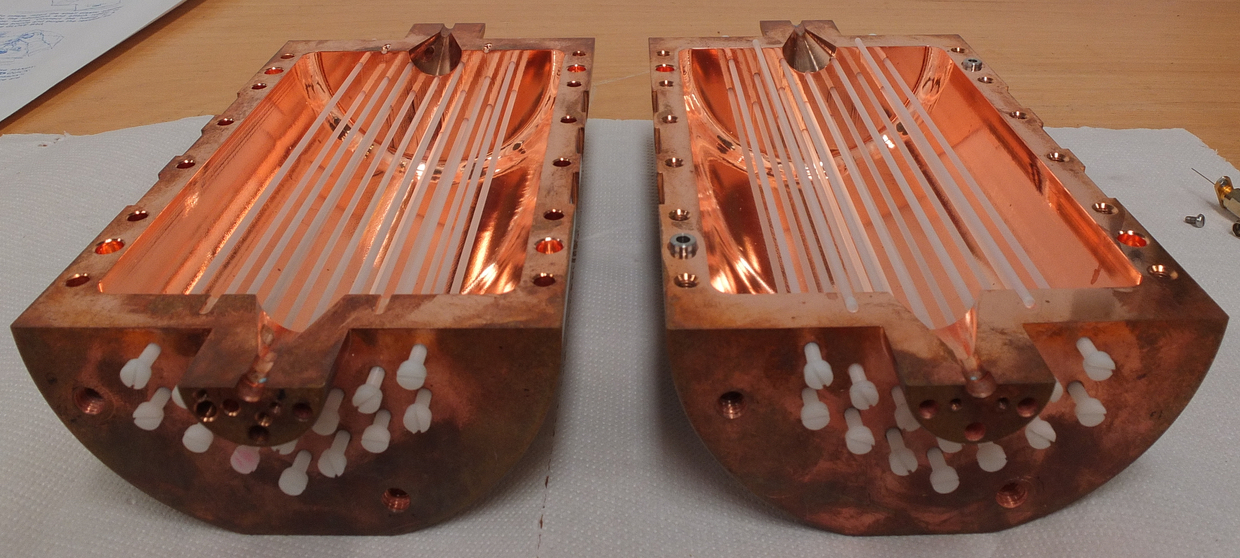}%
\caption{\label{cavity} Photograph of the  cavity opened in two halves.}%
\end{figure}

To realize the cavity two pieces of oxygen free high purity copper were machined to form two separate identical halves of the cavity. The inner surfaces were electro polished to reduce surface roughness. The two parts can then be joined together to form a closed cylinder, tightness is guaranteed by twelve M4 steel bolts. Bolts have been chosen of non magnetic iron (AISI 304). Before closing the cavity, the 36 sapphire rods are mounted. For each of them a 2.1 mm diameter hole is present on each ends of the two cavity halves.  Excluding the two most central ones, all sapphire rods are kept in place by teflon screws provided with copper berillium springs at their ends. The use of the springs is necessary to avoid breaking of the rods when cooling the system to cryogenic temperatures and, at the same time, keeping a constant force on the rods to avoid unwanted displacements. 
Figure \ref{cavity} shows the two halves of the cavity, complete of sapphire rods and teflon screws.

\section{Measurements}

\begin{figure}[h]
\includegraphics[width=.45\textwidth]{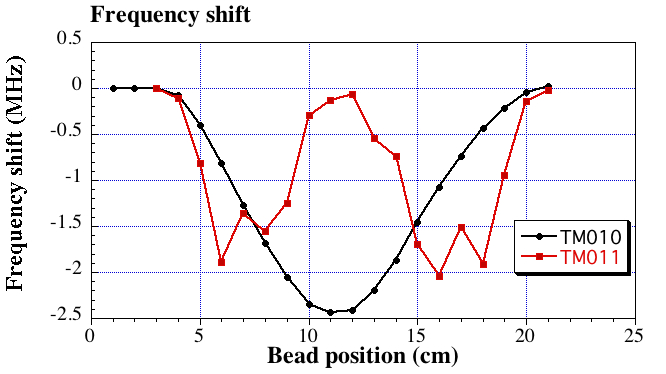}
\includegraphics[width=.45\textwidth]{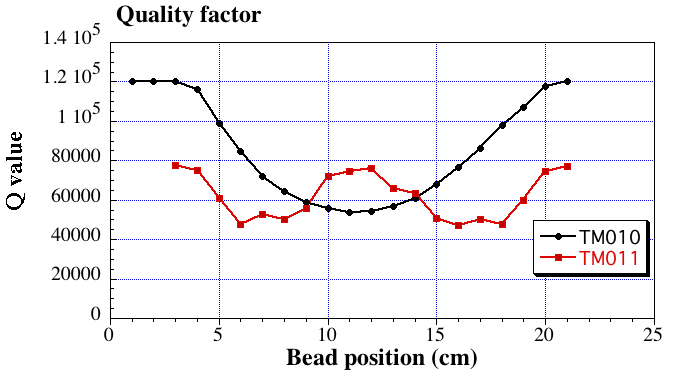}%
\caption{\label{bead} Bead pulling results (see text).}%
\end{figure}

The first measurement of the cavity is the identification of the principal modes. This has been performed, at room temperature, by using the standard bead pulling method, where a small dissipative sphere is moved along the cavity axis while the quality factor and frequency of selected modes are monitored. These parameters are obtained by measuring the S12 power transmission spectrum using a Vector Network Analyser (VNA), connected to  input/output ports on the cavity. Since the standard antennas are located on the cavity axis, two other holes were drilled just slightly off axis to be used as input/output ports, while the standard antenna holes where used as passages for a thin wire holding the bead. 
In figure \ref{bead} the results of these measurements are shown. The TM010 mode is identified with a single maximum for both quality factor and frequency shift with respect to a scan of the entire cavity length with the bead. The TM011 shows two maxima.  The measured frequency for the two modes were also compatible with the calculated ones, thus permitting a clear identification of the modes.
During this measurements care was taken to have both I/O ports with small coupling, so that no correction has to be taken into account due to the antennas.

Once the modes were identified, antennas have been mounted in the standard central holes along the cavity axis. The length of the dipole antennas has been chosen in such a way to have minimum coupling, but still measurable with the VNA, so that the  quality factor directly measured by the VNA practically coincides with the cavity unloaded $Q_0$. 


The cavity has been placed inside a vacuum chamber designed to allow operation inside cryogenic dewars. 
The vacuum chamber is equipped with two rf feedthroughs and a thermometer measures the temperature of the cavity. 
 
Measurements have been performed at room temperature, liquid nitrogen  and liquid helium temperature. Before cooling, the vacuum chamber is filled with about 10 mbar of helium to speed up the cooling process of the cavity. Low temperature rf measurements have been performed when the cavity temperature have reached a stationary value.

The starting point, with the cavity at room temperature, is as follows: mode TM010: both antennas  coupled at  -0.04 dB, resonance frequency $f_{\rm TM010} = 13 \,468$ MHz, linewidth $\Delta f_{\rm TM010} = 76$ kHz, unloaded quality factor $Q_{0,\rm TM010} = 173\, 000$.

During the cooling phase, due to the changes in the dimensions of the cavity, the cavity modes change their frequency accordingly. Moreover, probably due to changes in the relative positions of the sapphire rods, new modes show up and interfere with the main modes, thus making  the identification of the modes at low temperature difficult. For this reason, during the cooling the drift of the principal modes was constantly followed by taking continuos S12 spectra. 
\begin{figure}[h]
\includegraphics[width=.45\textwidth]{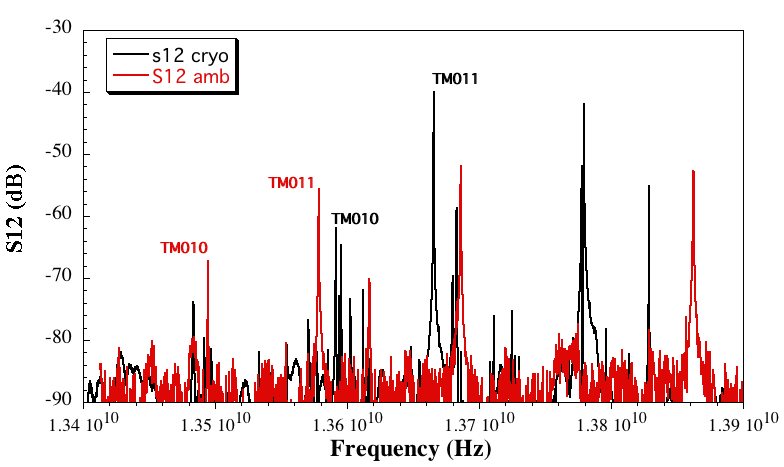}
\caption{\label{largesspectra} S12 Spectra at room (red curve) and cryogenic (black curve) temperature ($T=5.5$ K).}%
\end{figure}

Figure \ref{largesspectra} show the S12 spectra for room temperature and liquid helium temperature measurements. On the spectra the identified modes have been labelled. 
It has to be noted that the number and position of the unwanted modes are different for different cooling process. We are now working to make this phenomena reproducible.

\begin{figure}[h]
\includegraphics[width=.4\textwidth]{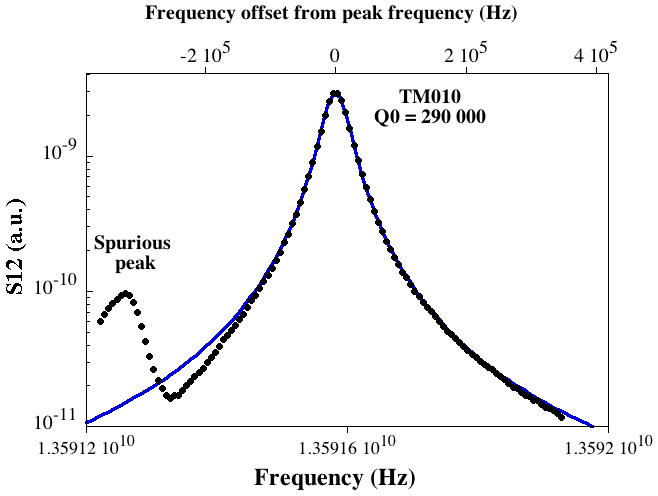}
\caption{\label{TM010} Measured S12 spectra (black dots) of the TM010 mode at liquid Helium temperature, superimposed with a fit (blue line) made with a Lorentzian function.}%
\end{figure}

To get a correct value of the quality factor, a zoomed spectra of the TM010 mode has been measured. This is then fitted with a Lorentzian line shape. Results are showed in figure \ref{TM010}. It can be seen that this mode is not anymore clear but another mode is interfering with it, thus reducing the quality factor to a lower value than the one expected from simulations. 

Table \ref{misure} summarizes the main results obtained. 
\begin{table}[ht]
\caption{Results of the measurements on the cavity.}
\begin{center}
\begin{tabular}{|c|c|c|c|}
\hline
System & Cavity Temperature & TM010  $Q_0$  & TM011 $Q_0$ \\ 
\hline
Room T & 298 K & 173 000 & 94 000 \\
\hline
Liquid He & 5.5 K & 290 000 & 520 000 \\
\hline
\end{tabular}
\end{center}
\label{misure}
\end{table}%


\section{Conclusions}

In this paper we have presented a novel type of cavity suitable to be  operated in a strong magnetic field. These cavities are ideal devices in the search for dark matter axions, where normal conducting ones are normally used. With the photonic cavity, an order of magnitude improvement of the quality factor is expected, while keeping the same detection volume. In fact, a cylindrical cavity working at the same frequency would have a cross  section of about 17 mm diameter, corresponding to an area of $2.3 \times 10^{-4}$ m$^2$.
Taking the same length of our photonic cavity, the effective volume is $V_{\rm eff}^{\rm Cu} = C_{mnl} V = 0.69 \times 3.4 \times 10^{-5}= 2.5 \times 10^{-5}$ m$^3$, to be compared with the value $V^{\rm Sa}_{\rm eff} = 2.3 \cdot 10^{-5}$ m$^3$ given in Section \ref{design}. This shows that the two cavities are equivalent for the purpose of axion searches. It is then possible to calculate what is the output power expected by using a set-up with, for example, an 8 T magnetic field. We use this number since in our laboratory tests have been already performed with a recently available superconducting magnet.

The expected axion power would be:
\begin{eqnarray}\nonumber
P_{\rm ax} = 3.3 \cdot 10^{-24}\, {\rm W} \left( \frac{V^{\rm Sa}_{\rm eff}}{2.3 \times 10^{-5}\, {\rm m}^3} \right)
\left( \frac{B}{8\, {\rm T}} \right)^2 \times\\
\left( \frac{g_\gamma}{-0.97} \right)^2
\left( \frac{\rho_a}{0.45\, {\rm GeV\, cm}^{-3}} \right)\nonumber
\left( \frac{f}{13.5\, {\rm GHz}} \right)
\left( \frac{Q_L}{145\,000} \right)
\end{eqnarray}

that could be reached with an almost quantum limited detection system integrating for a time span

\begin{eqnarray}\nonumber
{\rm t_m} = \frac{k_B^2 T^2_{\rm n}}{P^2_{\rm ax}} \frac{f}{Q_a} \simeq 2.4\cdot 10^5\,  {\rm s}\,  
\left( \frac{T_{\rm n}}{1\, {\rm K}} \right)^2 \times\\
\left( \frac{3.3 \cdot 10^{-24}\, {\rm W}}{P_{\rm ax}} \right)^2\nonumber
\left( \frac{f}{13.5\, {\rm GHz}} \right)
\left( \frac{10^6}{Q_a} \right),
\end{eqnarray}

where $k_B = 1.38 \cdot 10^{-23} $ J K$^{-1}$ is the Boltzmann constant, $T_{\rm n}$ is the system noise temperature and $Q_a$ the axion linewidth.
A stable operation of a Josephson Parametric Amplifier with a system noise temperature around 1 K has been already obtained at a slight different frequency in our laboratory at the Laboratori Nazionali di Legnaro of INFN \cite{Crescini20}.

We are thus currently working to implement this new cavity type in our haloscope, and meanwhile we are also studying other sapphire geometries that could boost the quality factor to even higher values.

\section*{Acknowledgements}

We are very grateful for the work of M. Zago that realized the technical drawings of the cavity. 
E. Berto, A. Benato and M. Rebeschini executed the mechanical work, while F. Calaon and M. Tessaro helped with the electronics and cryogenics, and F. Stivanello helped with the chemical works.
We acknowledge the work of the Cryogenic Service of the Laboratori Nazionali di Legnaro, providing us with large quantities of Liquid Helium on demand.


%
%

%


\bibliography{2019FotonicaV1}

\end{document}